\documentclass[prl,twocolumn,superscriptaddress,floatfix]{revtex4}

\usepackage{epsfig,dcolumn,amsmath,latexsym}

\usepackage{subfigure}
\usepackage[normalem]{ulem}
\usepackage{cancel}
\usepackage{multirow}

\begin{document}

\title{Pressure Dependent Electronic Structure in CeRh$_6$Ge$_4$} \preprint{1}

\author{Chao Cao}
 \email[E-mail address: ]{ccao@hznu.edu.cn}
 \affiliation{Condensed Matter Group,
  Department of Physics, Hangzhou Normal University, Hangzhou 310036, P. R. China}
 \affiliation{Center for Correlated Matter, Zhejiang University, Hangzhou 310058, China}

\author{Jian-Xin Zhu}
 \email[E-mail address: ]{jxzhu@lanl.gov}
 \affiliation{Theoretical Division and Center for Integrated Nanotechnologies, Los Alamos National Laboratory, Los Alamos, New Mexico 87545, USA}
\date{\today}

\begin{abstract}
 Using the state-of-art dynamical mean-field theory combined with density functional theory method, we have performed systematic study on the temperature and pressure dependent electronic structure of ferromagnetic quantum critical material candidate CeRh$_6$Ge$_4$. At -3.9 GPa and -8.3 GPa, the Ce-4$f$ occupation variation, the local magnetic susceptibility, and the low-frequency electronic self-energy behaviors suggest the Ce-4$f$ electrons are in the localized state; whereas at 6.5 GPa and 13.1 GPa, these quantities indicate the Ce-4$f$ electrons are in the itinerant state. The characteristic temperatures associated with the coherent Kondo screening is gradually suppressed to 0 around 0.8 GPa upon releasing external pressure, indicative of a local quantum critical point. Interestingly, the momentum-resolved spectrum function shows that even at the localized state side, highly anisotropic $\mathbf{k}$-dependent hybridization between Ce-4$f$ and conduction electrons is still present along $\Gamma$-A, causing hybridization gap in between. The calculations predict 8 Fermi surface sheets at the local-moment side and 6 sheets at the Kondo coherent state. Finally, the self-energy at 0.8 GPa can be well fitted by marginal Fermi-liquid form, giving rise to a linearly temperature dependent resistivity.
\end{abstract}

\maketitle

The electronic structure of strongly correlated system is subtly dependent on external stimuli. Application of pressure, magnetic field or chemical doping can tune competing ground states, giving rise to quantum phase transition (QPT)~\cite{Sachdev475,Coleman:2005aa,Sachdev:2008aa}. Although QPT is defined at 0K, the competition between the ground states invalidates the quasi-particle picture at finite temperatures close to the transition point, leading to quantum critical behavior. In addition, it is conjectured that new phases including superconductivity may emerge around the quantum critical point, as a result of the strong quantum fluctuations~\cite{RevModPhys.81.1551}. 

Quantum critical behaviors in Kondo lattice system were classified into two categories. In these systems, the Kondo coupling competes with the Ruderman-Kittel-Kasuya-Yosida (RKKY) interaction~\cite{Gegenwart:2008aa}. The former leads to a screening of local moments, while the later aligns the magnetic moments. Therefore, if the full screening remains at the onset of the local-moment alignment, the QPT is spin-density-wave (SDW) like, as described by the Hertz-Millis-Moriya theory~\cite{PhysRevB.14.1165,PhysRevB.48.7183,Moriya_1985_book}. At sufficiently low temperatures, the electronic structure at both sides of magnetic QCP are Fermi-liquid (FL) like. However, if the screening is insufficient, a local QCP exists, where the full screening can be achieved only at one side of QCP~\cite{Si:2001aa}. In this case, the electronic structure experiences a small to large fermi-surface (FS) transition across the QCP. For ferromagnetic (FM) QCP, it was proposed within the SDW picture that only the first-order transition or incommensurate magnetic ordering is allowed~\cite{PhysRevLett.92.147003}. Therefore, only a local QCP picture is possible for a uniform FM case, although it was argued that strong spin-orbit coupling and absence of inversion symmetry may complicate the argument~\cite{PhysRevLett.124.147201}. Recently, the discovery of CeRh$_6$Ge$_4$ has provided one of the rare examples with possible FM QCP in clean system~\cite{Shen:2020aa}, because unlike YbNi$_4$(P$_{1-x}$As$_x$)$_2$~\cite{Steppke933} and SrCo$_2$(Ge$_{1-x}$P$_x$)$_2$~\cite{Jia:2011aa}, CeRh$_6$Ge$_4$ can be tuned using pressure only. 
In this Letter, we unravel for the first time the nature of quantum phase transition in this system by using a  first-principles correlated electron approach.

In order to study the electronic structure of CeRh$_6$Ge$_4$, we performed DFT+DMFT calculations. Details of the calculation parameters can be found in the Supplementary Information (SI). The pressure effect was studied by expanding and shrinking the experimental lattice constant while fixing the internal atomic coordinates. In order to obtain a self-consistent description of electronic structure and external pressure and compare to experiments, we fitted the DMFT total energies at 232 K to the Birch-M\"{u}rnaghan equation of state (Tab. \ref{tab:ev}). At each pressure, the temperature dependence of electronic structure is studied using 13 different temperatures from 12 K to 580 K and two additional temperature (6 K and 9 K) for $-3.9$ GPa to 6.5 GPa.

\begin{table}
  \caption{Total energy-volume relation from calculations. $V_0$ is the calculated unit-cell volume at ambient-pressure (176.84 \AA$^3$). $E_{\mathrm{tot}}$ is the DMFT total energy obtained at 232 K ($\beta=50$ eV$^{-1}$), relative to the ambient pressure result. The corresponding pressure $p$ is obtained from Birch-M\"{u}rnaghan equation of state, with fitted bulk modulus of 172.5 GPa.  \label{tab:ev}}
  \begin{tabular}{c|c|c|c|c|c|c}
  $V/V_0$ & 0.882 & 0.937 & 0.966 & 0.995 & 1.024 & 1.055 \\
    \hline\hline
  $E_{\mathrm{tot}}$ (Ry) & 0.1251 & 0.0317 & 0.0089 & 0.0009 & 0.0045 & 0.0201 \\
  \hline
  $p$ (GPa) & 29.3 & 13.1 & 6.5 & 0.8 & $-3.9$ & $-8.3$ \\
  \hline\hline  
  \end{tabular}
\end{table}

\begin{figure*}
  \includegraphics[width=18cm]{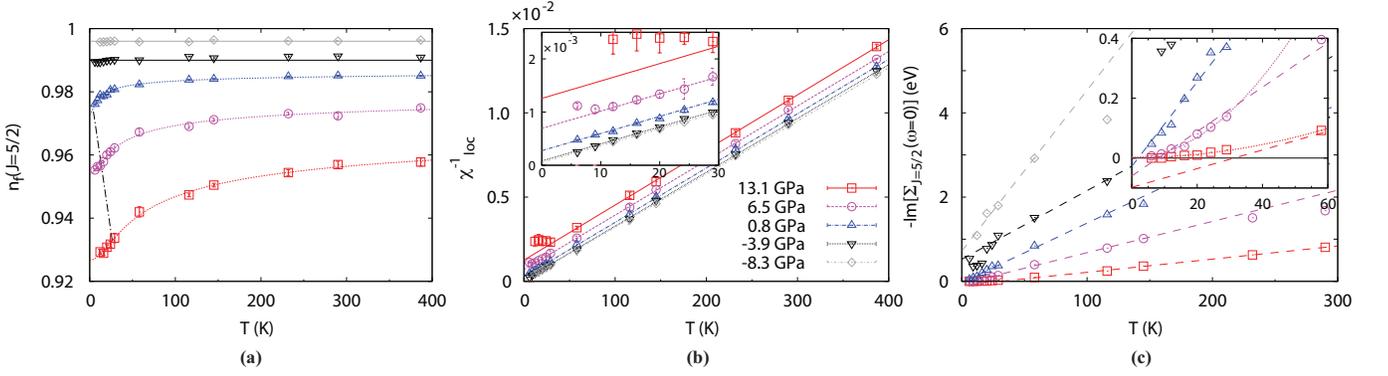}
  \caption{The temperature dependence of (a) occupation of Ce-4$f_{\mathrm{J}=5/2}$ states, (b) the inverse local magnetic susceptibility $1/\chi_{\mathrm{loc}}$, and (c) imaginary part of zero-frequency self-energy $\mathrm{Im}[\Sigma(\omega=0)]$. In each panel, the grey diamonds are for -8.3 GPa results, black down triangles -3.9 GPa, up blue triangles 0.8 GPa, pink circles 6.5 GPa, and red squares 13.1 GPa.  In panel (a) and (b), the statistical errors calculated using last 5 iterations of CT-QMC are shown as well. In panel (a), the dashed lines are fittings using the semi-empirical function as explained in the text, while the solid lines are eye candies. The dash-dotted line intersects the dashed lines at their respective $T_c$. In panel (b), the lines are high-temperature fittings between 58 K and 580 K. In panel (c), the dashed lines are high temperature linear fittings, while the dotted lines in the inset are low-temperature fittings to the Fermi-liquid form $\Sigma(0)\propto\alpha T^2$.
 \label{fig:dmft_stats}}
\end{figure*}

We start our discussion by examining negative pressure state far from the QCP, namely at $-8.3$ GPa. At $-8.3$ GPa, the system exhibits apparent local-moment behavior. The variation of Ce-4$f_{\mathrm{J}=5/2}$ occupation (between 0.995 at 12 K to 0.996 at 387 K) is negligibly small [Fig. \ref{fig:dmft_stats}(a)], suggesting insufficient Kondo screening of the local moment by the conduction electrons. It is consistent with the temperature dependence of the local magnetic susceptibility $\chi_{\mathrm{loc}}$, whose inverse can be well fitted to a linear function $a_{\chi}T+b_{\chi}$ of the temperature $T$ down to the lowest simulated temperature 12 K [Fig. \ref{fig:dmft_stats}(b)]. The fitted $b_{\chi}$ is negligibly small (order of 10$^{-5}$), with the coefficient of determination $R^2$=99.98\%. We must point out that if the system enters an ordered magnetic phase, the magnetic susceptibility must saturate and deviate from this line~\cite{PhysRevLett.91.156404,PhysRevLett.99.227204}. Because the Curie temperature for this system is too low to simulate in our calculation,  the saturation is not observed in our calculation. Nevertheless, it is worthy noting that in LDA+$U$ calculations, the FM phase is the ground state, with its total energy $\sim$130 meV lower than that of the NM phase, while AFM phase cannot be stabilized. Furthermore, its imaginary part of zero-frequency self-energy $\Sigma_0=\mathrm{Im}[\Sigma(\omega=0)]$ at low temperatures (below 145 K) is also linearly dependent on the temperature $T$ with its extrapolation to approximately $-0.8$ eV at 0 K [Fig. \ref{fig:dmft_stats}(c)]. Therefore, we conclude that the Ce-4$f$ electrons are fully localized, and these observations are consistent with the Kondo-break-down scenario at $-8.3$ GPa.

\begin{figure*}
  \subfigure[]{\includegraphics[width=8cm]{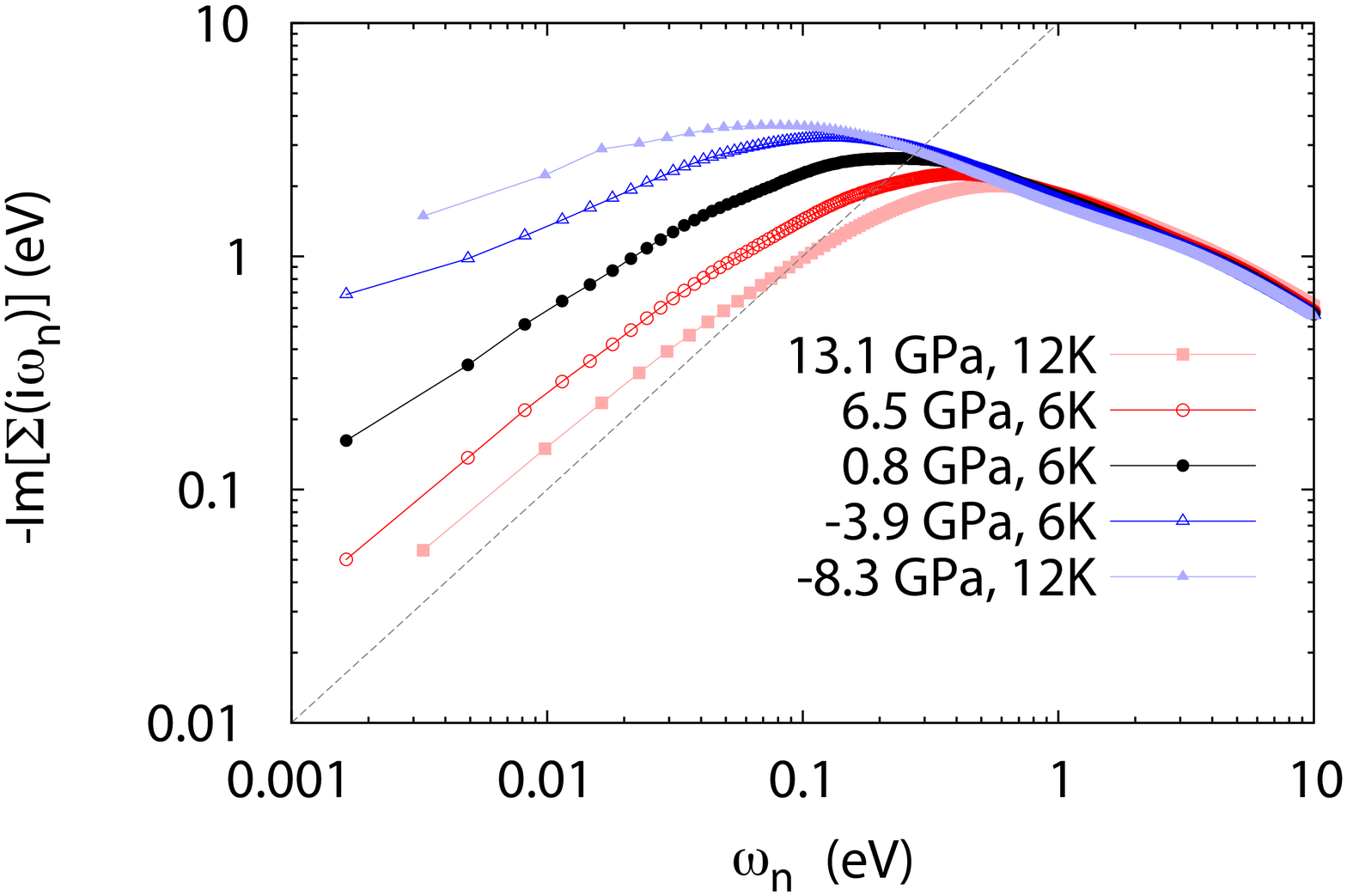}}
  \subfigure[]{\includegraphics[width=8cm]{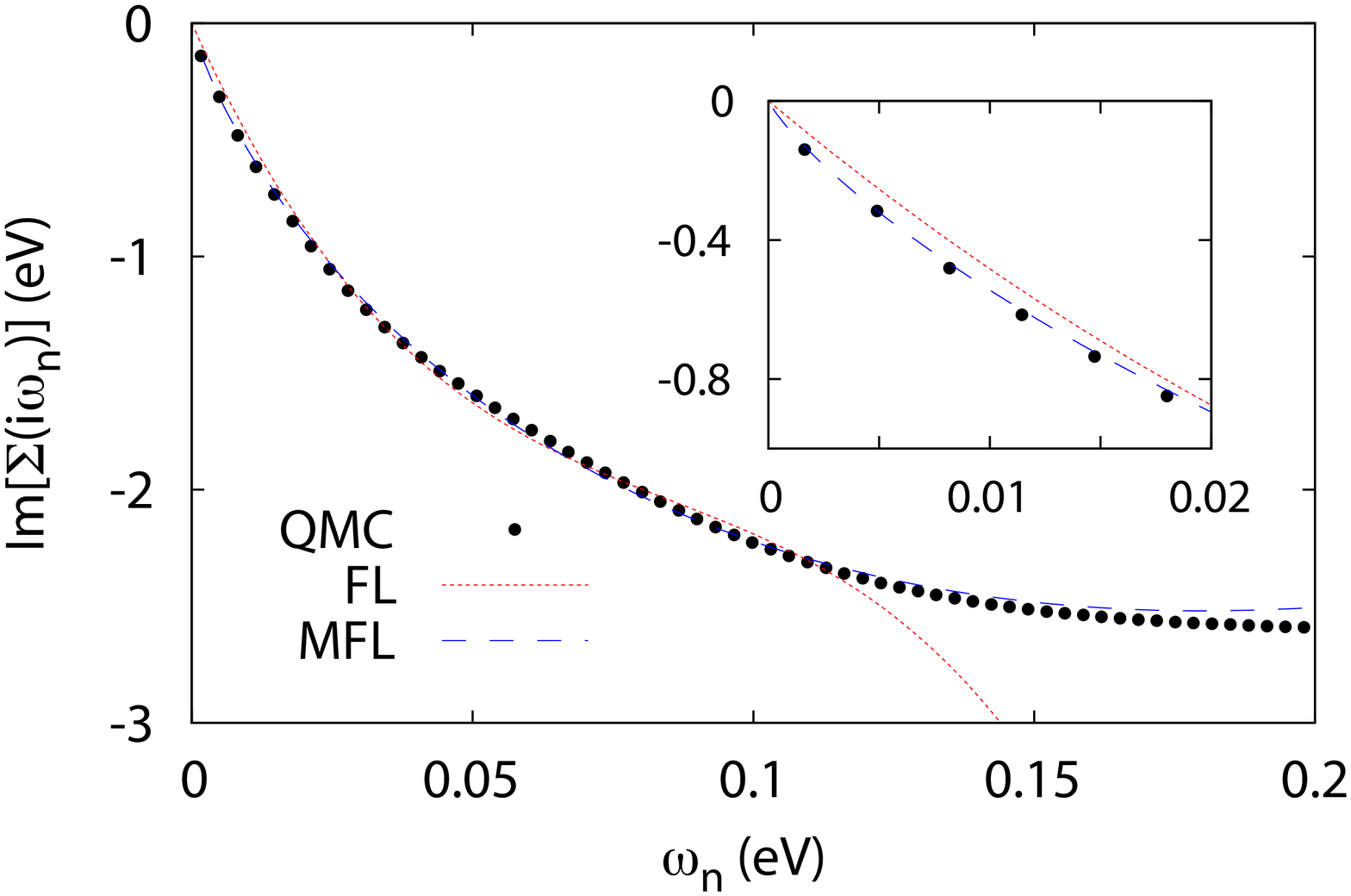}}
  \caption{The imaginary part of self-energy for Ce-4f$_{J=5/2}$ states in Matsubara frequencies. (a) Data obtained at the lowest simulated temperature at different pressure in logarithm scales and (b) data at 6K 0.8 GPa, with fitting to marginal Fermi-liquid form $\mathrm{Im}[\Sigma(i\omega_n)]\approx a\omega\log(\omega/b)$ and cubic polynomial form $a+b\omega+c\omega^2+d\omega^3$. The later is consistent with Fermi-liquid form at sufficiently low frequencies. \label{fig:sigma}}
\end{figure*}

The local-moment behavior at $-8.3$ GPa is fundamentally different from the results obtained at 13.1 GPa. At 13.1 GPa, the Ce-4$f_{\mathrm{J}=5/2}$ occupation $n_f(\mathrm{J}=5/2)$ varies from 0.958 at 387 K to 0.929 at 12 K, or equivalently more than 3\% change [Fig. \ref{fig:dmft_stats}(a)]. The temperature dependence of $\Delta n_f(T)$ can be fitted with a semi-empirical function $\Delta n_f(T)=\Delta n_0/[1+(T/T_c)^2]^{0.2}$~\cite{PhysRevLett.81.5225}. The fitted $T_c=26.8\pm6.6$ K (the uncertainty is chosen to be of 95\% fidelity) is a characteristic temperature that is associated with valence change. The inverse of the magnetic susceptibility $1/\chi_{\mathrm{loc}}$ clearly deviates from linear behavior below a characteristic temperature $T^S=$29 K, although the Curie-Weiss behavior persists in the high temperature range [Fig. \ref{fig:dmft_stats}(b)]. Such deviation is a clear indication of local-moment interacting with conduction electrons, suggesting formation of coherent Kondo screening. In addition, the imaginary part of zero-frequency self-energy $\Sigma_0(T)=\mathrm{Im}[\Sigma(\omega=0)]$ between 58 K and 290 K can be fitted to a linear function $a_{\Sigma}T+b_{\Sigma}$ with $b_{\Sigma}=0.098$ eV and coefficient of determination $R^2=$99.87\% [Fig. \ref{fig:dmft_stats} (c)]. Due to the causality requirement, positive $b_{\Sigma}$ suggests that $\Sigma_0(T)$ must deviate from the linear behavior and approaches 0. Therefore, the intersect of high-temperature linear fit and the $T$-axis defines another characteristic temperature $T^{\mathrm{FL}}=31.3\pm3.92$ K, which can be approximately assigned with Fermi liquid behavior. Indeed, $\Sigma_0(T)$ below 58 K can be well fitted to a parabolic function with coefficient of determination $R^2$=99.7\%. We also fit the low-frequency behavior of self-energy $\mathrm{Im}[\Sigma(i\omega_n)]$ at 13.1 GPa with Fermi-Liquid form $\Sigma_0(T)+(1-Z^{-1})\omega_n$. The obtained $\Sigma_0(T)$ below 58 K are diminishingly small (order of 0.01 eV), and the effective electron mass $m^{*}/m=Z^{-1}$ increases from 9.3 at 290 K to 18.5 at 58 K, and saturates around $m^*/m=17$ below 58 K. The formation of heavy Fermi-liquid behavior at low temperatures thus unambiguously puts the Ce-4$f$ electrons  at 13.1 GPa at the itinerant side.

Having established the Ce-4$f$ electrons in CeRh$_6$Ge$_4$ are localized at $-8.3$ GPa and itinerant at 13.1 GPa, we now analyze the states closer to the experimentally observed QCP, i.e. $-3.9$ GPa, 0.8 GPa, and 6.5 GPa. The temperature dependence of Ce-4$f_{\mathrm{J}=5/2}$ state occupation $n_f(\mathrm{J}=5/2)$ becomes gradually more evident with increasing pressure [Fig. \ref{fig:dmft_stats}(a)]. The variation of $n_f(\mathrm{J}=5/2)$ from 387 K to 12 K is less than 0.1\% (0.991 to 0.990) at -3.9 GPa, approximately 0.6\% (0.984 to 0.978) at 0.8 GPa, and about 2\% (0.975 to 0.956) at 6.5 GPa. At 6.5 GPa, we can also fit the variation of Ce-4$f_{\mathrm{J}=5/2}$ occupation $n_f(\mathrm{J}=5/2)$ to the same empirical function $\Delta n_f(T)=\Delta n_0/[1+(T/T_c)^2]^{0.2}$ with $T_c=13.09\pm3.9$ K. At 0.8 GPa, such fitting is marginal, because the uncertainty ($\sim$4 K) is larger than the characteristic $T_c=3.8$ K. Therefore, $T_c$ appears to be gradually suppressed around 0.8 GPa, which clearly indicates increasing tendency of Ce-4$f$ electrons being localized with decreasing pressure.

\begin{figure*}
  \includegraphics[width=18cm]{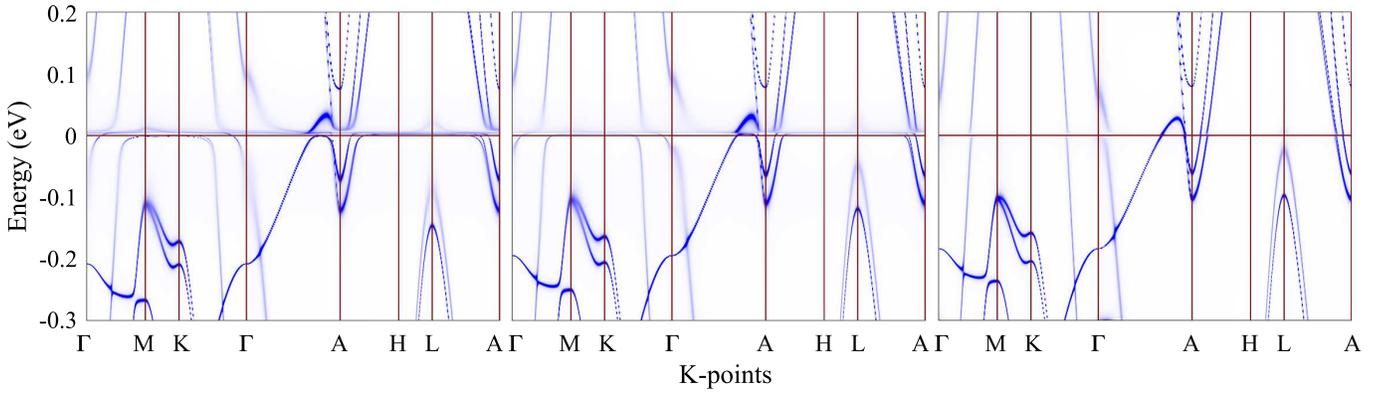}
  \caption{The momentum resolved spectrum function $A(\omega, \mathbf{k})$ at 6K for 6.5 GPa (left panel), 0.8 GPa (middle panel), and -3.9 GPa (right panel).  \label{fig:spectrum}}
\end{figure*}

The inverse magnetic susceptibility $1/\chi_{\mathrm{loc}}$ also shows systematic behavior [Fig. \ref{fig:dmft_stats}(b)]. At high temperatures, $\chi_{\mathrm{loc}}$ can always be fitted to a Curie-Weiss form $\chi_{\mathrm{loc}}^{-1}=a_{\chi}T+b_{\chi}$. At $-3.9$ GPa, the Curie-Weiss behavior extends to the lowest simulated temperature (6K), and the fitting almost overlaps with the $-8.3$ GPa data. In $-3.9$ and $-8.3$ GPa cases, both fittings yield negligibly small $b_{\chi}\sim10^{-5}$. At 0.8 GPa, no apparent deviation from the linear fitting is observed down to 6 K, and $b_{\chi}$ is $2.74\times10^{-4}$. We remark that $b_{\chi}$ should be even smaller if the spatial fluctuation is correctly accounted for~\cite{PhysRevLett.92.147003}. At higher pressures, the fitted $b$ significantly enhances as the pressure increases, and $b_{\chi}$ at 6.5 and 13.1 GPa is about 1 order of magnitude higher than that at 0.8 GPa and 2 orders of magnitude higher than that for  $-3.9$ and $-8.3$ GPa. In addition, the 6.5 GPa data begin to deviate from its high-temperature fitting below 12 K, and is clearly off the fitted line at 6 K. Therefore,  the characteristic temperature $T^S$ is $\approx 9\pm3$K at 6.5 GPa, and are gradually suppressed to 0 for pressures lower than 0.8 GPa. 

In Fig. \ref{fig:dmft_stats}(c), we show the pressure-dependent imaginary-part of zero-frequency self-energy $\Sigma_0(T)$ plot. At higher temperatures, $\Sigma_0$ is also linearly dependent on $T$, and are therefore can be fitted with a linear function $a_{\Sigma}T+b_{\Sigma}$ [the dashed lines in Fig. \ref{fig:dmft_stats}(c)]. For 0.8 GPa and 6.5 GPa, the fitted lines intersect with the $T$-axis at $T^{\mathrm{FL}}=1.95\pm1.05$K and $8.05\pm3.29$ K, respectively. For $-3.9$ GPa and $-8.3$ GPa, both fittings intersect with $T$-axis at $T<0$. Similar to our previous discussion, the characteristic temperature $T^{\mathrm{FL}}$ is also gradually suppressed to 0 around 0.8 GPa. In addition to this systematic change in $T^{\mathrm{FL}}$, $b_{\Sigma}$ also changes monotonically from negative to positive values with increasing pressure. For 6.5 GPa, deviation from the linear behavior is guaranteed by the causality requirement, and can be observed below 20 K. However, for 0.8 GPa, $b_{\Sigma}$ is nearly 0, suggesting that the system is very close to QCP at 0.8 GPa, and the linear behavior persists down to the lowest temperature for which we simulated (6 K). Since the imaginary-part of zero-frequency self-energy $-\mathrm{Im}[\Sigma(\omega=0)]$ represents the inverse life-time of quasi-particles, or the scattering rate, which can be approximated as proportional to the resistivity at low temperatures. Therefore, the linear behavior of $-\mathrm{Im}[\Sigma(\omega=0)]$ at 0.8 GPa is in line with the strange-metal behavior observed in transport experiment. 

We now focus on  self-energies at the lowest temperature in our simulations. In the Fermi-Liquid theory, at zero-temperature limit, the low-frequency self-energy obeys $\mathrm{Im}[\Sigma(i\omega_n)] \propto \omega_n$. For Ce-4$f$ in the paramagnetic local-moment state, since the scattering due to fluctuating local moment is finite at 0 K, the leading term of $\mathrm{Im}[\Sigma(i\omega_n)]$ should be a finite constant at small frequencies.  Therefore, it is expected that, in between,  the leading term of the self-energy $\mathrm{Im}[\Sigma(i\omega_n)] \propto \omega_n^{\alpha}$ with $\alpha<1$ around the local-itinerant type QCP. We plot the lowest temperature self-energies in logarithmic scales [Fig. \ref{fig:sigma}(a)]. For 13.1 GPa and 6.5 GPa, the logarithmic plot exhibits linear behavior at low-frequencies with their slopes close to 1, manifesting their Fermi-Liquid behavior. For $-3.9$ GPa and $-8.3$ GPa, the low-frequency behavior appears to approach a flat line in the logarithmic plot, consistent with the local-moment behavior with a leading constant term. For 0.8 GPa, the low-frequency logarithmic plot also exhibits a linear behavior, but its slope $\alpha$ is 0.68 rather than 1. In addition, for $T=6$ K, $\mathrm{Im}[\Sigma(i\omega_n)]$ at 0.8 GPa can be nicely fitted to the marginal Fermi-liquid (MFL) form $\mathrm{Im}[\Sigma(i\omega_n)]\approx a\omega\log(\omega/b)$ [Fig. \ref{fig:sigma}(b)], which can yield linearly temperature dependent resistivity~\cite{PhysRevLett.63.1996}. This analysis is also consistent with the temperature dependent effective mass $m*/m$ calculation (refer to SI). At 6.5 GPa, the effective mass increases from 10.1 at 145 K to 27.5 at 9 K with decreasing temperature, but it saturates below 9 K. At 0.8 GPa, $m*/m$ increases from 9.6 at 145 K to 60.2 at 6 K, without a clear saturation. Therefore, the effective mass shall diverge around the QCP. It is worthy noting that in either MFL form or $\omega_n^{\alpha}$ ($\alpha<1$) expansion, the diverging effective mass is naturally expected since $\lim_{\omega_n\rightarrow0} \frac{\partial \mathrm{Im}[\Sigma(i\omega_n)]}{\partial \omega_n}\rightarrow\infty$.

\begin{figure}
  \includegraphics[width=8.5cm]{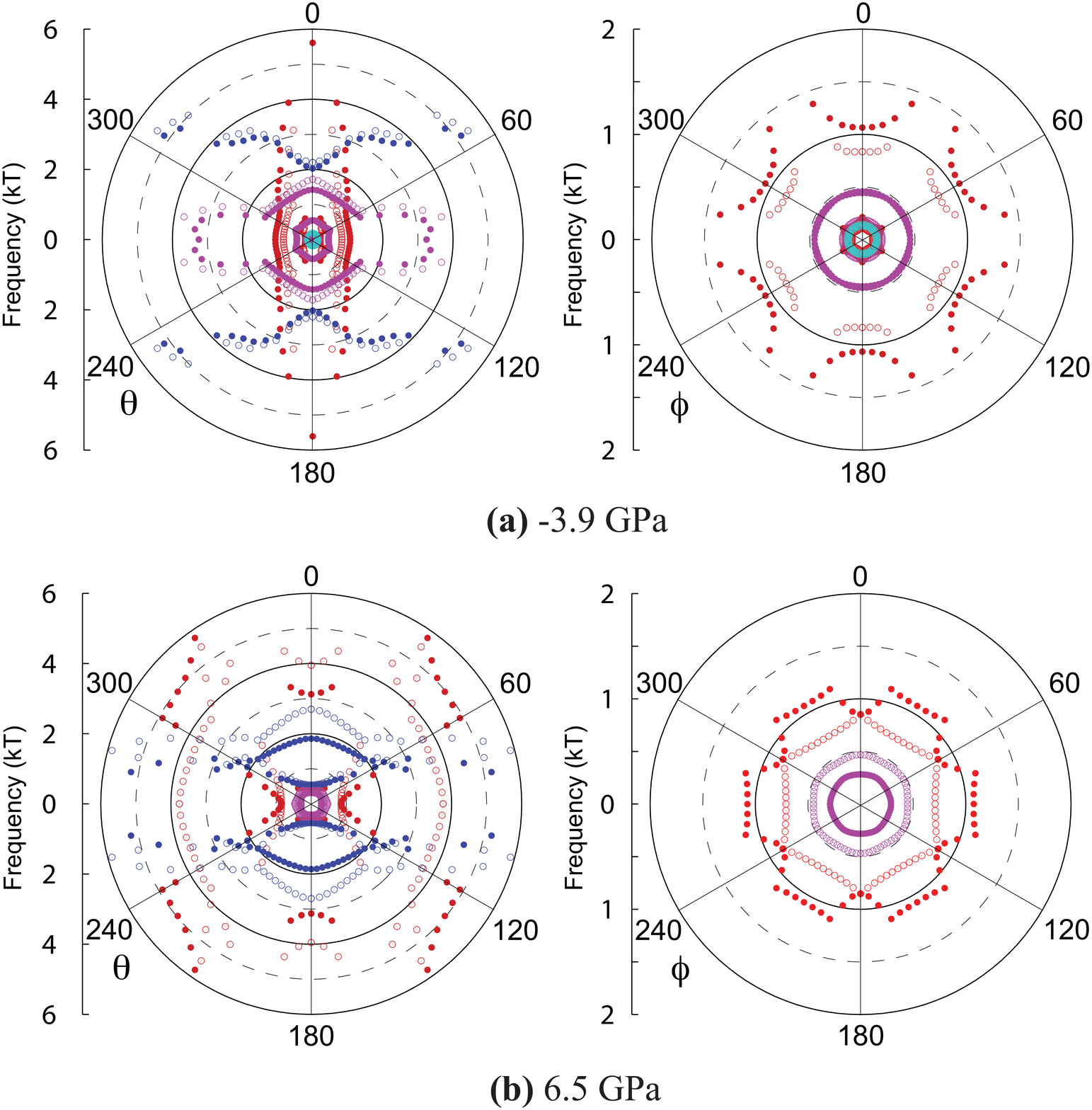}
  \caption{Polar plot of the angular dependent dHvA frequencies for (a) local-moment side (-3.9 GPa) and (b) itinerant side (6.5 GPa). In each case, the left panels are dependence on the polar angle $\theta$, measured with respect to [001]-direction within the (1$\bar{1}$0) plane; whereas the right panels are dependence on the azimuthal angle $\phi$, measured with respect to [100]-axis within the (001) plane. \label{fig:dhva}}
\end{figure}

We now show in Fig. \ref{fig:spectrum} the momentum-resolved spectral function of CeRh$_6$Ge$_4$ at 6 K for various values of pressure. At 6.5 GPa, a clear coherent flat Ce-4$f$ band can be identified near the Fermi level, suggesting the formation of full Kondo screening. In this case, the Fermi surface consists of 6 sheets, resembles the result from density-functional calculation assuming itinerant Ce-4$f$ electrons in the valence band. At -3.9 GPa, it is interesting to notice a clear indication of ligand conduction band bending between $\Gamma$-A, and formation of a fairly large hybridization gap. Such a band bending or hybridization gap is absent in the A-H-L plane, and is much smaller in the $\Gamma$-M-K plane. This strongly suggests  anisotropic hybridization between the Ce-4$f$ orbitals and ligand conduction bands. Nevertheless, flat bands due to Ce-4$f$ cannot be observed near the Fermi level, suggesting that the Kondo screening is only partial and dynamic instead of coherent~\cite{SI200623}. In this case, the Fermi surface of CeRh$_6$Ge$_4$ consists of 8 sheets, with their shape and size similar to DFT calculations assuming Ce-4$f$ electrons in core-state. In addition, the Ce-4$f$ electrons interact most strongly with two small Fermi surface pockets around A, consistent with the FM ground state obtained in experiment and in LDA+$U$ calculations. We have also calculated the angle-dependent dHvA frequencies in both cases (Fig. \ref{fig:dhva}). The calculated frequencies at $-3.9$ GPa reasonably agree with the experimental observations at ambient pressure~\footnote{Private communication with Huiqiu Yuan, manuscript submitted.}, suggesting that the Ce-4$f$ electrons are localized at ambient pressure. Our prediction on the electronic structure and Fermi surface topology at high pressure should be accessible to experimental verification.

In conclusion, we have performed a systematic DFT+DMFT study on the electronic structure of CeRh$_6$Ge$_4$ at different pressure and temperatures. The local magnetic susceptibility, Ce-4$f$ occupation, and low-frequency self-energy behavior suggest that the Ce-4$f$ form local moments at $-8.3$ and $-3.9$ GPa, and are itinerant at 6.5 GPa and 13.1 GPa. We have further predicted that  at the critical pressure of 0.8 GPa, the system exhibits marginal Fermi-liquid  behavior. In addition, highly anisotropic $f$-conduction hybridization occurs even in local moment regime. The Fermi surface consists of 8 sheets in the local-moment region and 6 sheets in the itinerant region, and the ambient pressure measurement agrees reasonably well with the local-moment calculation. We note that although the current calculation ignores spatial quantum fluctuations, our calculation yields local-QCP near experimental pressure (0.8 GPa), suggesting that local fluctuation dominates the quantum phase transition in CeRh$_6$Ge$_4$.

\begin{acknowledgements}
The authors are grateful to Jianhui Dai, Yang Liu, Michael Smidman and Huiqiu Yuan for stimulating discussions. This work was support by NSFC 11874137 (C.C.), LANL LDRD Program (J.-X.Z.), and in part by Center for Integrated Nanotechnologies, a DOE BES user facility. The calculations were performed on clusters at the High Performance Computing Cluster of Center of Correlated Matters at Zhejiang University, High Performance Computing Center at Hangzhou Normal University, and Tianhe-2 Supercomputing Center.
\end{acknowledgements}

\bibliography{ce164}

\end{document}